\documentclass[twocolumn,pre,showpacs,floatfix]{revtex4}
\usepackage{graphicx}
\usepackage{amsfonts}
\usepackage{amssymb}
\usepackage{bm}

\newcommand{\be}{\begin{equation}}
\newcommand{\ee}{\end{equation}}
\newcommand{\bea}{\begin{eqnarray}}
\newcommand{\eea}{\end{eqnarray}}
\newcommand{\zncu}{ZnCu$_3$(OH)$_6$Cl$_2$}

\begin{document}

\title{Triplet and Singlet Excitations in the
Valence Bond Crystal Phase of Kagome Lattice Heisenberg Model}

\author{Rajiv R.~P.~Singh}
\affiliation{Department of Physics, University of California, Davis,
CA 95616, USA}

\author{David A.~Huse}
\affiliation{Department of Physics, Princeton University, Princeton,
NJ 08544, USA}
\date{\today}


\begin{abstract}
A proposed ground state for the Kagome Lattice Heisenberg Model
(KLHM) consists of a Valence Bond Crystal (VBC) with a 36-site unit
cell. We calculate the low-lying triplet and singlet excitations in
the VBC phase for the infinite-lattice model and
for the 36-site cluster. 
For the infinite lattice, the lowest triplet excitation is found to
have a spin gap of approximately $0.08 \pm 0.02 J$ and a bandwidth
of only about $0.01 J$. For the 36-site cluster, consisting of a
single unit cell with periodic boundary conditions, there are
substantial finite-size effects: the spin gap there is estimated to
be approximately $0.2 J$, close to the exact diagonalization result
of $0.164 J$. The triplet excitations attract one another and form
many bound states in the spin-singlet channel.  We find a large
number of such bound states for the 36-site cluster, many of which
appear to lie below the spin gap, again in agreement with the
results from exact diagonalization.

\end{abstract}

\maketitle

\section{Introduction}

Ground state properties of the Kagome-Lattice Heisenberg Model have
been studied by a wide variety of numerical and analytical
techniques
\cite{review,elser,zeng90,singh92,leung93,elstner,lhuillier,waldtmann,mila,misguich05,MS,
marston,nikolic,maleyev,auerbach,hermele05,ran06}. Very recently, we
used \cite{singh-huse} series expansion methods to make the case
that a Valence Bond Crystal (VBC) phase consisting of a honeycomb
lattice of `perfect hexagons' with a 36-site unit cell (shown in
Fig.~1) is the correct ground state phase of this model. We found
the ground state energy of this state to be $-0.433\pm 0.001 J$,
which is lower than other existing variational estimates. We also
showed that with periodic boundary conditions on the 36-site cluster
the energy of the state is further lowered by an amount that makes
it compatible with the result from exact diagonalization.

\begin{figure}[!htb]
\vspace{0.5cm}
\begin{center}
  \includegraphics[scale=.5]{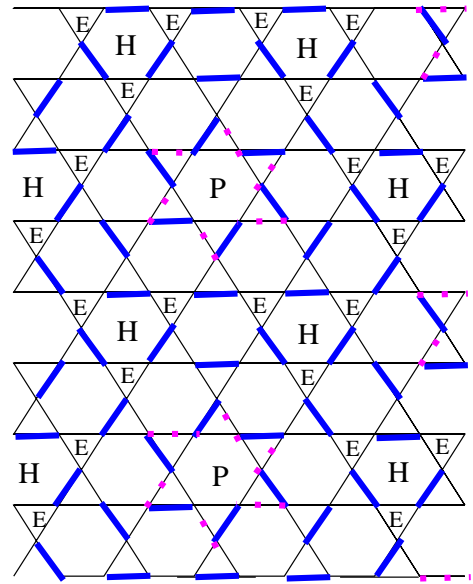}
\end{center}
\vspace{-0.2cm} \caption{\label{pattern} (Color online) Ground state
ordering pattern of low-energy (``strong'') bonds (blue) for the
Kagome Lattice Heisenberg Model. The perfect hexagons are denoted as
H, the empty triangles by E and the pinwheels as P. The two dimer
coverings of the pinwheels that remain degenerate to high orders of
perturbation theory are denoted by thick solid (blue) and dotted
(magneta) bonds.}
\end{figure}

One of the most intriguing features of the model is the finding of a
large number of singlet  states below the lowest triplet in the
exact diagonalization studies of finite clusters
\cite{lhuillier,waldtmann,mila,misguich05,MS}.  The VBC phase, with
a large unit cell, provides a natural starting point for explaining
these states. In the 36-site cluster, the broken symmetries 
of the VBC phase would give rise to 48 low-lying singlet states for
every singlet state in the broken-symmetry phase. However, there are
several respects in which such an explanation has so far been
unsatisfactory. First, the exact diagonalization studies do not see
a clear separation between a `tower' of ground states, and the other
singlet states; the spectrum appears to be a continuum. Second, if a
triplet is the lowest excitation in the VBC phase, one would expect
only 48 singlets below the lowest triplet, whereas the number of
such states is found to be about 200. Third, the quantum numbers of
the lowest-lying states of the 36-site cluster do not match with the
quantum numbers of the states 
that arise from linear combinations of VBC ground states \cite{MS}.

The fact that the $36$-site cluster consists of only a single unit
cell of the VBC allows substantial finite-size effects and a strong
overlap of the various ground and excited state manifolds. This was
evident from our earlier study, where we found that various states
of the thermodynamic system have energy differences of order $0.001
J$ per site, whereas the periodic boundary condition on the 36-site
cluster leads to changes in energy of order $0.006 J$. This means
that there is unlikely to be a definitive signature of the broken
symmetry in the spectra of this cluster. It is the purpose of this
paper to understand the excitation spectrum of the VBC phase,
especially to identify the singlets that fall within the spin gap
and to explore their relation to the states obtained in the exact
diagonalization study.

The 36-site cluster is large enough that we apply the same
linked-cluster methods to develop series expansion for the finite
cluster that we used to study the infinite system. Thus both for the
infinite system and for the 36-site cluster, we have obtained the
triplet spectra correct to $7$-th order in the dimer expansion.
These series are not long enough to capture the superposition of
different ground and excited states in the 36-site cluster, which
restore all symmetries of the (undimerized) kagome lattice and lead
to states with well defined crystal momenta and point group quantum
numbers in the full Brillouin zone. That would require a separate
diagonalization study of the model in a restricted basis set
\cite{MS}, which is not part of the calculations presented here.
Still, as will see, the series expansion study of triplet and
singlet excitations proves quite informative.

The VBC unit cell has 18 dimers. Thus the triplet spectrum is
obtained from the eigenvalues of an $18\times 18$ matrix. For the
infinite system, this matrix depends on the momentum in the reduced
Brillouin zone of the VBC.  Note that all momenta $q$ of the 36-site
cluster map to $q=0$ in the reduced Brillouin zone. The series
coefficients are much more convergent for the infinite system than
for the finite cluster, consistent with what is expected if the true
ground state of the infinite system is a VBC. We estimate the spin
gap to be $0.08\pm 0.02 J$ for the infinite system, with the lowest
band of triplets having its minimum at $q=0$.  Our series estimates
are less accurate for the $36$-site cluster, where we obtain a spin
gap $\Delta$ of roughly $0.22 J$. The exact diagonalization result
for the triplet gap is $0.16419 J$ \cite{sindzingre2000}. This
difference is within the uncertainties of our calculations of this
quantity on the finite cluster.

The two-particle calculations are more difficult, and we have only
carried out a general calculation on the 36-site cluster and only to
second order; we find as many as $15$ different singlet bound states
of two triplets.  This order is not enough to accurately estimate
the absolute energy of these bound states. Just adding up the terms
gives a negative energy, in other words, the excitation would fall
below the ground state. However, this is also the case for the
triplets.  For the triplets, where we have longer series, we know
that higher-order
terms produce positive excitation energies. 
The expansions for 
the binding energies appear more reasonable.  To second order, the
binding energies of the lowest four states, in units of J, are 0.33,
0.28, 0.25 and 0.25 respectively.  These lie between $\Delta$ and
$2\Delta$, implying that these singlet excitations lie within the
spin gap.  These states together with the ground state can provide
up to $5\times 48=240$ singlet states below the spin gap in the
36-site cluster, roughly consistent with exact diagonalization
results.

The plan of this paper is as follows. In section II, we outline the
series expansion methods used for the calculations. In section III,
we discuss the ground state properties, including the dimerization
order parameter. In section IV, we discuss the triplet excitations.
In section V, the singlet bound state calculations are discussed.
Finally, in section VI, we discuss in more detail the relationship
of our work to earlier exact diagonalization studies of the 36-site
cluster.

\section{Series expansions for the VBC phase}
We are interested in the properties of the Kagome Lattice Heisenberg
Model (KLHM) with Hamiltonian,
\begin{equation}
H=J\sum_{\langle i,j\rangle} {\bf S}_i\cdot{\bf S}_j,
\end{equation}
where $S_i$ are spin-half operators and the sum runs over all
nearest-neighbor bonds of the Kagome Lattice. In order to carry out
the series expansion study, we write the Hamiltonian as $H_0+\lambda
H_1$, where $H_0$ consists of all Heisenberg exchange terms on the
strong bonds of the VBC phase and $H_1$ consists of all other bonds.
Standard Raleigh-Schrodinger perturbation theory in powers of
$\lambda$ is used to study the ground state properties of the system
\cite{series-reviews}.  The series are extrapolated to $\lambda=1$
either by summing up terms in the series or by using Pade
approximants. To study the triplet excitation spectra, we use the
similarity transformation method, first discovered by Gelfand,
details for which can be found in the literature
\cite{series-reviews,book}.

For any finite (or infinite) Lattice, series expansion calculations
by the Linked Cluster method require an enumeration of all possible
graphs (or clusters) that can be embedded in that lattice up to some
size, and a calculation of various recursion relations on those
graphs or clusters. We find it convenient to define the graphs by
the set of strong bonds (or dimers) that are contained in them, and
all weak bonds connecting those strong bonds are included in a
single graph. In this sense, our graph enumeration scheme is similar
to a low temperature expansion (or strong embedding). The difference
between the infinite system and the $36$-site system with periodic
boundary conditions (PBC) arises from the different graphs that can
be embedded in them. In the $36$-site system there are only 18
dimers, whereas for the infinite lattice their number is infinite,
but graphs related by a translation of the unit cell are counted
only once. Since the system has a large unit cell, it is important
to distinguish between the different inequivalent bonds in the
lattice. We have used a simple counting scheme, where all connected
graphs of the $36$-site cluster are treated separately. For the
infinite system, all graphs that are distinct under the translation
symmetries of the VBC are treated separately. This leads to 19,837
graphs with up to 8 dimers on the 36-site cluster and 67,861 graphs
with up to 8 dimers for the infinite Kagome Lattice. While many of
these clusters are topologically equivalent, we have not exploited
their equivalence. Our calculations are limited by the counting
associated with the complex unit cell. These graphs are sufficient
for calculating ground state energy correct to $8$-th order and
other quantities to $7$-th order. Our results for the infinite
system and the 36-site cluster agree completely with our earlier
calculations of the ground state properties to $5$-th order and
triplet spectra to second order based on a very small number of
specially defined topological graphs, and the longer series continue
the trend of surprisingly good convergence for the infinite lattice
that we noted earlier \cite{singh-huse}.

\section{Ground State Properties: Dimerization Order Parameters}

In this section, we discuss the ground state energy and dimerization
order parameters for the Kagome Lattice Heisenberg Model. The ground
state energy series for the infinite lattice is found to be
\begin{widetext}
\be E_0^{(\infty)}/J=-{3\over 8} -{3\over 64}\lambda^2 -{1\over
256}\lambda^3 -0.00577799\lambda^4 -0.000528971\lambda^5
-0.000634087\lambda^6
 +0.0000952526 \lambda^7 -0.0000470303 \lambda^8.
\ee
\end{widetext}
This leads to an estimate for the ground state energy of $-0.4326\pm
0.0001 J$. All error bars in this paper are subjective estimates of
the uncertainties in series extrapolation to $\lambda=1$. They do
not reflect any statistical uncertainties in the calculations. For
quantities where we can simply sum up the terms in the series, we
estimate their value by their partial sum up to the smallest term in
the series (in magnitude) and set the uncertainty to be roughly
twice the magnitude of the smallest term.  We have checked that the
numbers are consistent with Pade approximant estimates.

The dimerization order parameters are defined by the expectation
values $\langle {\bf S}_i \cdot {\bf S}_j\rangle$ on the different
bonds. Note that the ground state energy implies a mean energy per
bond of approximately $-0.216 J$. In the large unit cell, there are
many inequivalent bonds.  We focus here on the dimerization order
parameters inside the `perfect hexagons'. If we consider an isolated
hexagon and write our Hamiltonian as $H_0+\lambda H_1$, with $H_0$
containing the exchange on the strong bonds, the symmetry will be
restored as $\lambda$ goes to unity and the dimerization order
parameter will be zero. In the infinite Kagome Lattice, either the
dimerization order parameter inside the `perfect hexagons' can go to
zero, or the coupling with the rest of the lattice (and other
hexagons) can generate a self-consistent dimerization in each
hexagon. We find here that the latter case is realized.  For the
infinite lattice, the series for the expectation value on the strong
bond of the `perfect hexagons' (Bond Type A) is given by
\begin{widetext}
\begin{equation}
E_A=-0.75 +0.1875 \lambda^2  +0.046875\lambda^3  +0.0789388
\lambda^4 +0.00832791 \lambda^5  +0.00481033 \lambda^6
-0.0104363\lambda^7 -0.00798403 \lambda^8
\end{equation}
whereas the series for the weak bond in the perfect hexagons (Bond
Type B) is
\begin{equation}
E_B=-0.1875\lambda -0.0703125\lambda^2 -0.0683594\lambda^3
-0.0106608 \lambda^4 +0.00889079 \lambda^5  +0.0172855 \lambda^6.
\end{equation}
\end{widetext}
Note that the more difficult off-diagonal expectation values ($E_B$)
are carried out up to only $7$-dimer graphs. Summing to the smallest
term, the energy of bond type A is $E_A=-0.42\pm 0.01 J$, and bond
type B is $ E_B=-0.33\pm 0.02 J$; the difference is larger if we sum
the full series.  Note that they are both lower than the mean energy
per bond, and that there is a clear alternation inside the perfect
hexagons as well, indicating that nonzero VBC order parameters,
including dimerization within the perfect hexagons, survive to
$\lambda=1$.

It is interesting to compare these results with the Kagome-stripe
model, which was studied by DMRG \cite{white-singh}. The latter
consists of a single-hexagon wide strip of the Kagome Lattice. That
model was found to be spontaneously dimerized in a pattern
reminiscent of the `perfect hexagons' in the VBC phase of the KLHM.
One important difference is that in the latter model, the three-fold
rotational symmetry of the Kagome Lattice is absent and the
dimerization preferably occurs, not surprisingly, along the stripe
axis.

\begin{widetext}
We have also extended the ground state energy series for the 36-site
cluster. We obtain \be E_0^{(36)}/J=-{3\over 8} -{3\over
64}\lambda^2 -{1\over 256}\lambda^3 -0.00821940\lambda^4
-0.00224474\lambda^5 -0.000257987\lambda^6
 +0.00104012 \lambda^7 +0.00161714 \lambda^8.
\ee
\end{widetext}
Summing up to the smallest term leads to the estimate $-0.4365 J$;
Pade approximants show large uncertainties due to the strong growth
of the last two terms. This should be compared with the exact
diagonalization result of $-0.438377 J$. It is likely that a large
part of the difference is associated with the superposition of
different symmetry-broken ground states, an effect which is not
captured in our calculations, although it may be the cause of the
growth of the last two coefficients.

\section{Triplet Excitation Spectra}
In this section, we study the triplet excitation spectra. As
discussed in our earlier paper, an important aspect of the dimer
states of the KLHM is the existence of empty triangles, in which
there are no strong bonds. All quantum fluctuations originate from
the empty triangles. The empty triangles also play a key role in
studying the triplet excitations. There are 18 dimers in a unit cell
and thus at $\lambda=0$ there are 18 degenerate triplet excitations
for every $q$ in the reduced Brillouin zone, all with energy $J$.
These triplets can be classified into Light and Heavy particles.
Those dimers that border the empty triangles are much more mobile
and will be called Light triplets. These are the triplets that make
up the lowest energy excitations. There are nine such states per
unit cell, residing on the three dimers in each of the two perfect
hexagons and on the three dimers that connect two perfect hexagons
via empty triangles (the Light Bridging Dimers).  A numbering
scheme, together with the sign convention for the singlet states on
these dimers is shown in Fig.~2.  The other nine states are on the 6
pinwheel dimers and on the three dimers parallel to the Light
Bridging Dimers (which we call Heavy Bridging Dimers) and lead to
heavy or nearly immobile triplets.  The latter only indirectly
influence the low energy triplets, and that in a very marginal way.
Thus, we will focus our discussion here on the 9 Light Dimers. In
studying the singlet bound states in the next section, the Heavy
Bridging Dimers (see Fig.~2) will also play an important role.

\begin{figure}[!htb]
\vspace{0.5cm}
\begin{center}
  \includegraphics[scale=.5]{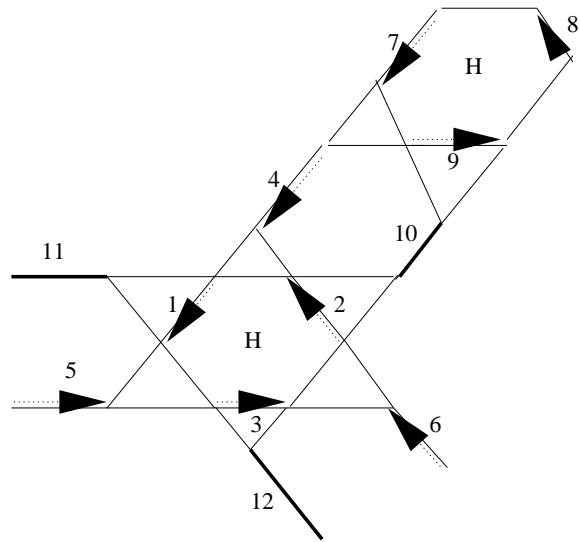}
\end{center}
\vspace{-0.2cm} \caption{\label{arrows} Labeling of the triplets in
a unit cell. The dimers in first perfect hexagon are labeled as 1, 2
and 3. The Light dimers that bridge the perfect hexagons (Light
Bridging Dimers L1, L2 and L3) are labeled as 4, 5 and 6. The dimers
in the second perfect hexagon are labeled 7, 8 and 9. The arrows
denote the convention for the singlet states on the bonds. The Heavy
dimers between the unit cells (Heavy Bridging Dimers H1, H2, and H3)
are shown by thick lines and labeled 10, 11 and 12. }
\end{figure}

The triplet spectra are studied by the similarity transformation
method, which integrates out the multiple triplet states, order by
order in perturbation theory, to generate an extended effective
Hamiltonian in the subspace of single triplets
\cite{series-reviews,book}. Our first task is to analytically study
the lifting of the degeneracy in the single triplet subspace. We
focus on the lowest lying triplet in the reduced Brillouin zone.
Once, the degeneracy has been lifted in some order of perturbation
theory, we simply need to carry out a non-degenerate perturbation
theory for that state. Let $z_1=\exp{i \vec k \cdot \vec r_1}$ with
$\vec r_1=-4 \sqrt{3} \hat y$, and $z_2= \exp{i \vec k \cdot \vec
r_2}$ with $\vec r_2=-6\hat x -2\sqrt{3} \hat y$, with the lattice
oriented as in Fig. 1, with a nearest-neighbor spacing of unity. In
the first order, the effective Hamiltonian is given by $({-\lambda
J/4}) M_1$, where the matrix $M_1$ is shown in Table I, where $z_3=
z_1^*z_2$. Note that the matrix depends on conventions of the unit
cell and of the choice of arrows, which define the singlet states on
the bonds. Note also that we have changed notation from our previous
paper \cite{singh-huse}, to make it more symmetric.

\begin{table}[h]
\caption{ Matrix $M_1$}
\begin{tabular}{rrrrrrrrr}
\hline\hline
  0&  -1&  -1&  -1&   1&   0&   0&   0&   0\\
 -1&   0&  -1&   1&   0&  -$z_1$&   0&   0&   0\\
 -1&  -1&   0&   0&  -1&   $z_1$&   0&   0&   0\\
 -1&   1&   0&   0&   0&   0&  -1&   0&   1\\
  1&   0&  -1&   0&   0&   0&   0&   $z_3$& $-z_3$\\
  0&  $-z_1^*$&   $z_1^*$&   0&   0&   0&   $1$& $-1$&  0\\
  0&   0&   0&  -1&   0&   $1$&  0&  -1&  -1\\
  0&   0&   0&   0&   $z_3^*$& -1& -1&  0&  -1\\
  0&   0&   0&   1&   $-z_3^*$& 0&   -1&  -1&  0\\
\hline\hline
\end{tabular}
\end{table}

For $q=0$ the matrix $M_1$ leads to 4 degenerate states. Two of
these are triplets localized within a `perfect hexagon'. These are
invariant under the $2 \pi/3$ rotational symmetry of the lattice,
the only symmetry that remains unbroken in the VBC.  The other two
states form a doublet that have a nonzero `angular momentum' under
this rotation.  By symmetry, the former pair of two states can not
mix with the latter. Higher order calculations show that the nonzero
angular momentum states have lower energy starting in 3rd order.
This is also what happens on the $36$-site cluster, which only has
$q=0$ in the reduced Brillouin zone.  The series for
the lowest triplet (the spin gap) on the infinite-lattice is 
\begin{widetext}
\begin{equation}
\Delta_\infty/J=
 1 -0.5\lambda -0.875\lambda^2  +0.440625\lambda^3
+0.074479167\lambda^4 -0.04346788\lambda^5 -0.02336301\lambda^6
-0.1487368 \lambda^7
\end{equation}
For the $36$-site cluster the series for the lowest triplet is
\begin{equation}
\Delta_{36}/J=
 1 -0.5\lambda -0.875\lambda^2  +0.440625\lambda^3
+ 0.4864583 \lambda^4 -0.1698405 \lambda^5 -0.4289909 \lambda^6
  +0.393891 \lambda^7
\end{equation}
\end{widetext}
Note that the difference first appears in 4th order, just as for the
ground state, as the wrapping around of the PBC begins to influence
the state. Clearly, the series for the infinite lattice is better
behaved, although the last term suggests that one does
not have absolute convergence.  
We can speculate that the lack of convergence might arise from the
process of lifting of the degeneracy in low orders and this can
have a `ringing' effect at higher orders. 
Using Pade approximants, we estimate the spin gap for the infinite
system to be $\Delta_\infty =0.08\pm 0.02 J$.

For the $36$-site cluster, Pade approximants lead to an estimate of
the gap of approximately $0.22 J$. This should be compared with the
exact-diagonalization result of $0.16419 J$. Given the irregularity
in this series, this is within the uncertainties of the calculation.
It is also likely that the superposition of different
symmetry-broken states that must appear in the finite-size cluster
at $\lambda=1$ may lead to a further lowering of the spin gap.  For
the ground state this extra energy is of order $0.002 J$ per lattice
site, so the total energy difference for the 36-site cluster is of
order $0.07 J$. One might expect energy corrections of this order
for the excitations as well.

For $q\ne 0$, there are only $3$ states that are degenerate at order
$\lambda$. Two of them are localized in the `perfect hexagons'. The
third is given by the vector in Table II, and is delocalized. Once
again, in third-order perturbation theory, the delocalized state
ends up with the lower energy.  On the infinite lattice, the
difference between the lowest energy excitation at any $q$ and that
at $q=0$ remains very small in all orders of the calculation that we
have studied so far.  We find a very small band width of order $0.01
J$, with the band minimum remaining at $q=0$.

\begin{table}[h]
\caption{ Nontrivial eigenvector of matrix $M_1$}
\begin{tabular}{r}
\hline\hline
$z_1+z_1z_3-2z_3$\\
$z_1z_3+z_3-2z_1$\\
$z_3+z_1-2z_1z_3$\\
$3(z_1-z_3)$\\
$3(z_3-z_1z_3)$\\
$3(z_3-1)$\\
$z_1+1-2z_3$\\
$z_3+z_1-2$\\
$1+z_3-2z_1$\\
\hline\hline
\end{tabular}
\end{table}

We note that in addition to the lowest-lying triplet band studied
here, there is a second low-energy band, which is more dispersive
and is degenerate with the lowest band at $q=0$. Then there are at
least two more low-lying triplet bands consisting of triplets which
are nearly localized inside a perfect hexagon and weakly hop from
one perfect hexagon to next. In contrast to the lowest triplet, the
series expansions for these latter states are poorly convergent for
the infinite lattice, perhaps reflecting the fact
that before we reach $\lambda=1$ they become unstable 
and can decay into the lower triplet and even lower-lying singlets.

\section{Singlet excitations}
In this section, we discuss the singlet bound states of two
triplets.  We focus only on the $36$-site cluster, which means that
the two triplets are restricted to a single unit cell with periodic
boundary conditions. 
The bound states can be divided into two types: (1) Light-Heavy
bound states and (2) Light-Light bound states.

\subsection{Light-Heavy Bound states}
The light-heavy bound states arise from the fact that when the heavy
triplet lies on one of the Heavy Bridging Dimers (10, 11 or 12 in
Fig.~2) and the light triplet lies on one of its neighboring dimers
in one of the `perfect hexagons' they have an attraction of
magnitude $-\lambda J/2$. However, in order $\lambda$ the light
triplet can also move around. Hence, to calculate the singlet bound
state and binding energy correct to order $\lambda$ we need to once
again consider a $9$-dimensional subspace, where the heavy triplet
is held fixed but the light triplet is allowed to hop between the
nine light dimers. This is very much like the matrix $M_1$. The only
difference is that when the light and heavy particles share a pair
of common weak bonds there is a diagonal binding energy of $-\lambda
J/2$. Diagonalizing the $9\times 9$ matrix, we find two bound
states.  In the more tightly bound state the wavefunction is
symmetric under the reflection that exchanges the two ends of the
occupied heavy dimer, whereas it is antisymmetric for the other
bound state.  Since the heavy triplet can be in 3 equivalent
positions, this implies a total of six light-heavy bound states on
the 36-site cluster.

To get the binding energies of the states to second order, we have
calculated the $9\times 9$ effective Hamiltonian to second order
given by
\begin{equation}
{\cal H}_{eff}= P H_1 (1-P) {1\over E_2 -H_0} (1-P) H_1 P,
\end{equation}
where $P$ is the projection operator on the two-particle subspace
and $E_2=2J$ is the bare energy of the two-particle states. 
Calculating the expectation
value of the effective Hamiltonian in the bound states obtained in
first order, leads to energies to second order for the bound states,
relative to the ground state,  of
\begin{equation}
E_{lh1}=2 -1.22545 \lambda -0.88485422 \lambda^2,
\end{equation}
and
\begin{equation}
E_{lh2}=2 -1.171040 \lambda -0.83436254 \lambda^2~.
\end{equation}
Note that if we set $\lambda$ to unity, this would mean a negative
energy for these states, i.e., an energy even below the ground
state. However, the second order calculation of the minimum triplet
energy is 
$1-0.5 \lambda-0.875 \lambda^2$. 
Thus, if we simply add up the terms at $\lambda=1$, these states do
not even remain bound.  Clearly a higher order calculation is needed
to get an estimate of the energy of these light-heavy states. That
is beyond the scope of the current work.


For the infinite system, the Light-Heavy bound state calculation at
order $\lambda$ is a tight-binding model for the light triplet with
a short-range impurity potential near the site of the heavy triplet.
The light triplet can go arbitrarily far away from the heavy
triplet, thus the calculation requires diagonalizing an
infinite-dimensional matrix already at order $\lambda$.  It seems
clear that the light-heavy bound state is nearly dispersionless. For
the state to develop any $q$-dependence the heavy triplet would have
to move from one unit cell to another and that is a very high order
process.

\subsection{Light-Light Bound states}
When two light triplets share a weak bond they also feel an
attraction in the singlet sector in order $\lambda$ of magnitude
$-\lambda J/4$.  To study the light-light bound state fully in order
$\lambda$, we need to consider the $C^9_2=36$-dimensional Hilbert
space in which the two light triplets occupy any two of the $9$
light dimers.  In this case, the diagonalization of the
$36$-dimensional matrix leads to $8$ different bound states.

We have also calculated the energy of these bound states to order
$\lambda^2$. The energies of various bound states to second order
are given in Table III. Note that at order $\lambda^2$ the
light-light bound states have significantly lower energies than the
light-heavy bound states.  If we simply add up the terms in the
series we will get strongly negative results, which are unphysical.
However, it is quite plausible that these large negative energies
simply follow from the large negative energies for individual
triplets at this order, which we now know are corrected at higher
orders.  Thus we look at the series for the binding energy by
subtracting at each order the energy of a bound state from the
lowest energy of two `free' triplets.

\begin{table}[h]
\caption{ Series Coefficients for the energies of the Light-Light
Bound states to order $\lambda^2$}
\begin{tabular}{rrrr}
\hline\hline
state& $\lambda^0$& $\lambda^1$&  $\lambda^2$\\
1  & 2 &-1 &-2.08333 \\
2  & 2 &-1.22554 &-1.81451 \\
3  & 2 &-1.08013  &-1.92010 \\
4  & 2 &-1.08013 & -1.92010\\
5  & 2 &-1.15062 & -1.71371 \\
6  & 2 &-1.15062 & -1.71371\\
7  & 2 &-1.14039 & -1.73862\\
8  & 2 &-1.14039 & -1.73862\\
9  & 2 &-1.14039 & -1.68347\\
\hline\hline
\end{tabular}
\end{table}

Simply evaluating the series at $\lambda=1$, we obtain binding
energies for these states of $0.33$, $0.29$, $0.25$, $0.25$, $0.13$,
$0.13$, $0.11$, $0.11$ and $-0.03$ times $J$, respectively.  So $8$
of the $9$ states are estimated to remain bound.  Furthermore, the
binding energy of the lowest four states exceeds the spin gap
$\Delta_{36}$ and this suggests that these singlet states will fall
below the lowest triplet.  While these second-order calculations
cannot be taken as quantitatively accurate estimates for the
energies of these singlet bound states, our calculations strongly
suggest that there are lots of singlet states which can fall within
the spin gap.  Since each state corresponds to $48$ distinct states
when the symmetry-related ground states are counted on the 36-site
cluster, these 4 singlet excitations together with the ground state
can lead to $48\times 5=240$ singlet states within the spin gap.
This is clearly a correct order of magnitude when compared to the
exact diagonalization.  The nature of the singlet bound states is
discussed in the next section.

For the infinite system, the light-light bound states can be
dispersive, as each of the triplets is dispersive.  However, given
that the light triplets have a very weak dispersion, the dispersion
of the bound state is likely to also be weak.  Our calculations,
thus, suggest that there are a few low-energy bands of one or more
weakly dispersive singlet excitations.  In the future, we hope to
develop systematic methods to study these singlet excitations and
their dispersion to higher order.

\section{Discussions: Nature of singlets and relation
to exact diagonalization studies} In this section, we first discuss
qualitatively the wavefunctions of the singlet bound states in the
order they are listed in Table III:

\subsection{Wavefunctions and symmetries of bound states}

1. The state with largest binding energy when truncated at order
$\lambda^2$ has a very simple wavefunction. One triplet is in one
perfect hexagon and the other triplet is in the other perfect
hexagon.  The two triplets attract each other by a very unusual
interference mechanism.  Since they are never on neighboring sites
they have no direct contact attraction and thus no binding at order
$\lambda$. The attraction arises from the fact that when the two
triplets are on dimers separated by a light bridging dimer that
shares a weak bond with both, then there is constructive
interference to go to a three-triplet state and back, and this
process significantly lowers the energy of this bound state in order
$\lambda^2$.

2. The second largest binding energy is for a state where both
triplets are mostly in the same perfect hexagon, and the state is
symmetric under the exchange of perfect hexagons.

3-6. A set of two separate doubly degenerate states (3,4 and 5,6 in
Table III) are the ones with nonzero `angular momentum'.  In these, there is
significant probability of finding one or both triplets to be
outside the perfect hexagons (on the light bridging dimers).  The
probability of finding both triplets in same perfect hexagon is very
small.  Since these states are bound states of what in third order
become the lowest-energy triplets, and as they will not have any
level repulsion from the ground state due to having different
symmetry, it seems quite possible that finally the lower of these
bound states end up with the lowest energy overall.

7-9. Finally there are a set of 3 states which are degenerate in
order $\lambda$ but split into a doublet and a singlet in order
$\lambda^2$. Their wavefunction has significant probability of
finding the two triplets in the same perfect hexagon, but the
wavefunction is antisymmetric under the exchange of the two
hexagons.

An important question is precisely how are the singlet bound states
studied here related to the low-lying singlet states found in the
exact diagonalization studies.  This issue cannot be quantitatively
addressed based on series expansion alone. What we cannot address
here is what happens when the 48 different ground states
corresponding to the symmetry-related dimer patterns are fully
admixed at $\lambda=1$. That is what restores the full symmetry of
the underlying kagome lattice, and leads to states which have
well-defined quantum numbers under all the symmetries of the
lattice.  To explore this issue further, it may be useful to study
the KLHM in a reduced basis set consisting of the various low-lying
states that we have found, and their symmetry partners.

We also note that low-lying singlets found in our study include some
with nonzero `angular momentum' and thus lower symmetry than the
honeycomb VBC, consistent with what is found in the exact
diagonalization studies.  It is also important to note that the
two-particle bound states generally lie outside the nearest-neighbor
singlet dimer subspace of the Heisenberg model.

\subsection{Relevance to Experiments}

Recently, there has been much interest in the material \zncu, as it
represents a first example of structurally perfect spin-half Kagome
Lattice \cite{helton07,ofer07,mendels07,imai07}.  Unfortunately,
this material may have significant anti-site disorder leading to
magnetic impurities and possibly also has significant
Dzyloshinski-Moria (DM) interactions \cite{rigol07,sindzingre07}.
Since it has a spin-gap in its spectrum, the VBC state might survive
a sufficiently small DM interaction.  Thus it is important to obtain
samples with reduced impurities and ascertain the strength of DM
interactions.

If the VBC phase is relevant to a material, Raman scattering can be
used to study the singlet states just as Neutron scattering is used
to study the triplet spectra. Even if there is short-range VBC
order, low lying singlet excitations may be sharply defined. Raman
scattering on \zncu \ should be used to look for low-lying singlet
states.

In the future, it would be useful to develop a more detailed
phenomenological picture of the experimental properties in the VBC
phase. That is distinct from the goals of the present work.

\begin{acknowledgments}

We would like to thank Philippe Sindzingre for several useful
communications. This work was supported by the US National Science
Foundation, Grants No.\ DMR-0240918 (R. S.) and DMR-0213706 (D. H.)
and PHY05-51164.

\end{acknowledgments}


\begin{thebibliography}{99}

\bibitem{review} For a recent review see C. Lhuillier, arXiv:cond-mat/0502464.

\bibitem{zeng90}
C. Zeng and V. Elser, Phys. Rev. B {\bf 42}, 8436 (1990).

\bibitem{elser} V. Elser, Phys. Rev. Lett. {\bf 62}, 2405 (1989).

\bibitem{singh92}
R. R. P. Singh and D. A. Huse, Phys. Rev. Lett. {\bf 68}, 1766
(1992).

\bibitem{leung93}
P. W. Leung and V. Elser, Phys. Rev. B {\bf 47}, 5459 (1993).

\bibitem{elstner} N. Elstner and A. P. Young
Phys. Rev. B {\bf 50}, 6871 (1994).

\bibitem{lhuillier} P. Lecheminant, B. Bernu, C. Lhuillier, L. Pierre and P.
Sindzingre, Phys. Rev. B {\bf 56}, 2521 (1997).

\bibitem{mila} F. Mila, Phys. Rev. Lett. {\bf 81}, 2356-2359 (1998).

\bibitem{waldtmann} C. Waldtmann, H. U. Everts, B. Bernu, C. Lhuillier,
P. Sindzingre, P. Lecheminant and L. Pierre,
 Eur. Phys. J. B {\bf 2}, 501 (1998).

\bibitem{misguich05}
G. Misguich and B. Bernu, Phys. Rev. B {\bf 71}, 014417 (2005).

\bibitem{MS} G. Misguich and P. Sindzingre,
J. Phys.: Condens. Matter {\bf 19}, 145202 (2007).

\bibitem{marston} J. B. Marston and C. Zeng, J. Appl. Phys. {\bf 69},
5962 (1991).

\bibitem{maleyev} A. V. Syromyatnikov and S. V. Maleyev, Phys. Rev. B{\bf 66},
132408 (2002).

\bibitem{nikolic} P. Nikolic and T. Senthil, Phys. Rev. B {\bf 68},
214415 (2003).

\bibitem{auerbach} R. Budnik and A. Auerbach, Phys. Rev. Lett. {\bf 93},
187205 (2004).

\bibitem{hermele05}
M. Hermele, T. Senthil, and M. P. A. Fisher, Phys. Rev. B {\bf 72},
104404 (2005).

\bibitem{ran06}
Y. Ran, M. Hermele, P. A. Lee, and X. -G. Wen, Phys. Rev. Lett. {\bf
98}, 117205 (2007).

\bibitem{singh-huse} R. R. P. Singh and D. A. Huse, Phys. Rev. B {\bf 76},
180407 (2007).

\bibitem{sindzingre2000} P. Sindzingre, G. Misguich, C. Lhuillier, B. Bernu,
L. Pierre, Ch. Waldtman, H-U Everts,
 {\it Physical Review Letters}, {\bf 84},  2953  (2000); P. Sindzingre
and C. Lhuillier, to be published.

\bibitem{series-reviews} M. P. Gelfand and R. R. P. Singh, Adv. Phys.
{\bf 49}, 93 (2000); M. P. Gelfand, R. R. P. Singh and D. A. Huse,
J. Stat. Phys. {\bf 59}, 1093 (1990).

\bibitem{book} J. Oitmaa, C. Hamer and W-H. Zheng, {\it Series Expansion
Methods for strongly interacting lattice models}, (Cambridge
University Press 2006).

\bibitem{white-singh} S. R. White and R. R. P. Singh
Phys. Rev. Lett. {\bf 85}, 3330 (2000).

\bibitem{helton07}
J. S. Helton, K. Matan, M. P. Shores, E. A. Nytko, B. M. Bartlett,
Y. Yoshida, Y. Takano, A. Suslov, Y. Qiu, J.-H. Chung, D. G. Nocera,
and Y. S. Lee,
Phys. Rev. Lett. {\bf 98}, 107204 (2007).

\bibitem{ofer07}
O. Ofer, A. Keren, E. A. Nytko, M. P. Shores, B. M. Bartlett, D. G.
Nocera, C. Baines, and A. Amato,
cond-mat/0610540.

\bibitem{mendels07}
P. Mendels, F. Bert, M. A. de Vries, A. Olariu, A. Harrison, F. Duc,
J. C. Trombe, J. Lord, A. Amato, and C. Baines,
Phys. Rev. Lett. {\bf 98}, 077204 (2007).

\bibitem{imai07}
T. Imai, E. A. Nytko, B. M. Bartlett, M. P. Shores, D. G. Nocera,
cond-mat/0703141.

\bibitem{rigol07}
M. Rigol and R. R. P. Singh, Phys. Rev. Lett. {\bf 98}, 207204
(2007).

\bibitem{sindzingre07} G. Misguich and P. Sindzingre, Eur. Phys. J. B {\bf 59},
305 (2007).

\end{thebibliography}
\end{document}